%
%
%
%
\input harvmac.tex


\def\np#1#2#3{Nucl. Phys. {\bf B#1} (#2) #3}
\def\pl#1#2#3{Phys. Lett. {\bf #1B} (#2) #3}
\def\prl#1#2#3{Phys. Rev. Lett. {\bf #1} (#2) #3}
\def\physrev#1#2#3{Phys. Rev. {\bf D#1} (#2) #3}

\def\cmp#1#2#3{Comm. Math. Phys. {\bf #1} (#2) #3}

\def\ap#1#2#3{Ann. Phys. {\bf #1} (#2) #3}

\nref\om{C. Montonen and D. Olive, ``Magnetic monopoles as gauge particles
?'', \pl {72}{1977}{117}; P. Goddard,
J. Nuyts, and D. Olive, ``Gauge theories and magnetic charge'', 
\np{125}{1977}{1}}%
\nref\osborn{H. Osborn, ``Topological charges for $N=4$
supersymmetric gauge theories and monopoles of spin 1'', \pl{83}{1979}{321}}%
\nref\hsw{P. Howe, K. Stelle and P. West, ``A class of finite
four-dimensional supersymmetric field theories'', \pl{124}{1983}{55}}%
\nref\swi{N. Seiberg and E. Witten, ``Monopole condensation and confinement
in $N=2$ supersymmetric Yang-Mills theory'', hep-th/9407087, 
\np{426}{1994}{19}}%
\nref\swii{N. Seiberg and E. Witten, ``Monopoles, duality and chiral
symmetry breaking in $N=2$ supersymmetric QCD'', hep-th/9408099, 
\np{431}{1994}{484}}%
\nref\seiberg{N. Seiberg, ``Electric-magnetic duality in supersymmetric
nonabelian gauge theories'', hep-th/9411149, \np{435}{1995}{129}}%
\nref\ls{R. G. Leigh and M. J. Strassler, ``Exactly marginal operators and
duality in four-dimensional $N=1$ supersymmetric gauge theory'',
hep-th/9503121, \np{447}{1995}{95}}%
\nref\bais{F. A. Bais, ``Charge-monopole duality in spontaneously broken
gauge theories'', \physrev{18}{1978}{1206}}%
\nref\dw{R. Donagi and E. Witten, ``Supersymmetric Yang-Mills theory and
integrable systems'', hep-th/9510101, IASSNS-HEP-95-78}%
\nref\vw{C. Vafa and E. Witten, ``A strong coupling test of S duality'',
hep-th/9408074, \np{431}{1994}{3}}%
\nref\ggpz{L. Girardello, A. Giveon, M. Porrati and A. Zaffaroni,
``Nonabelian strong-weak coupling duality in (string derived) $N=4$
supersymmetric Yang-Mills theories'', hep-th/9406128, \pl{334}{1994}{331};
``S duality in $N=4$ Yang-Mills theories with general gauge groups'',
hep-th/9502057, \np{448}{1995}{127}}%
\nref\sen{A. Sen, ``Dyon-monopole bound states, self-dual harmonic forms on
the multi-monopole moduli space, and $SL(2,Z)$ invariance in string
theory'', hep-th/9402032, \pl{329}{1994}{217}}%
\nref\porrati{M. Porrati, ``On the existence of states saturating the
Bogomol'nyi bound in $N=4$ supersymmetry'', hep-th/9505187, 
NYU-TH-95-05-03}%
\nref\ardo{P. C. Argyres and M. R. Douglas, ``New phenomena in $SU(3)$
supersymmetric gauge theory'', hep-th/9505062, \np{448}{1995}{93}}%
\nref\apsw{P. C. Argyres, M. R. Plesser, N. Seiberg and E. Witten, ``New
$N=2$ superconformal field theories in four dimensions'', hep-th/9511154,
RU-95-81,WIS-95-59,IASSNS-HEP-95/95}%
\nref\suspect{J. P. Gauntlett and J. A. Harvey, ``$S$-duality and the dyon
spectrum in $N=2$ super Yang-Mills theory'', hep-th/9508156, EFI-95-56; 
S. Sethi, M. Stern and E. Zaslow, ``Monopole and dyon bound states in $N=2$
supersymmetric Yang-Mills theories'', hep-th/9508117, HUTP-95-A031}%
\nref\ho{A. Hanany and Y. Oz, ``On the quantum moduli space of vacua of
$N=2$ supersymmetric $SU(N_c)$ theories'', hep-th/9505075, 
\np{452}{1995}{283}}%
\nref\aps{P. C. Argyres, M. R. Plesser and A. D. Shapere, ``The Coulomb
phase of $N=2$ supersymmetric QCD'', hep-th/9505100, \prl{75}{1995}{1699}}%
\nref\as{P. C. Argyres and A. D. Shapere, ``The vacuum structure of $N=2$
super-QCD with classical gauge groups'', hep-th/9509175, RU-95-61}%
\nref\hanany{A. Hanany, ``On the quantum moduli space of vacua of $N=2$
supersymmetric gauge theories'', hep-th/9509176, IASSNS-HEP-95-76}%
\nref\seibergp{N. Seiberg, ``Supersymmetry and nonperturbative beta 
functions'', \pl{206}{1988}{75}}%
\nref\mn{J. A. Minahan and D. Nemeschansky, ``Hyperelliptic curves for
supersymmetric Yang-Mills'', hep-th/9507032, USC-95-019}%
\nref\jumps{S. Cecotti, P. Fendley, K. Intriligator and C. Vafa, ``A new
supersymmetric index'', hep-th/9204102, \np{386}{1992}{405};
S. Cecotti and C. Vafa, ``On classification of $N=2$
supersymmetric theories'', hep-th/9211097, \cmp{158}{1993}{569}; M.
Henningson, ``Discontinuous BPS spectra in $N=2$ gauge theory'',
hep-th/9510138, YCTP-P14-95}%
\nref\weinberg{E. Weinberg, ``Fundamental monopoles and multimonopole
solutions for arbitrary simple gauge groups'', \np{167}{1980}{500}}%
\nref\weinbergg{E. Weinberg, ``Fundamental monopoles in theories with
arbitrary symmetry breaking'', \np{203}{1982}{445}}%
\nref\atiyah{M. F. Atiyah and N. Hitchin, {\it The Geometry and Dynamics of
Magnetic Monopoles}, Princeton University Press (1988)}%
\nref\zm{R. Jackiw and C. Rebbi, ``Solitons with fermion number $1/2$'', 
\physrev{13}{1976}{3398}}%
\nref\gauntlett{J. P. Gauntlett, ``Low energy dynamics of $N=2$
supersymmetric monopoles'', hep-th/9305068, \np{411}{1994}{443}}%
\nref\callias{C. Callias, ``Index theorems on open spaces'', 
\cmp{62}{1978}{213}}%
\nref\dyons{E. Witten, ``Dyons of charge $e \theta / 2 \pi$'',
\pl{86}{1979}{283}}%
\nref\ms{N. S. Manton and B. Schroers, ``Bundles over moduli spaces and the
quantization of BPS monopoles'', \ap{225}{1993}{290}}%
\nref\newmn{J. A. Minahan and D. Nemeschansky, ``$N=2$ super Yang-Mills and
subgroups of $SL(2,\bZ)$'', hep-th/9601059, USC-96-01}%

\def\ai{{\vec{\alpha}^i}}
\def\rp{{\rm Re}(\Phi)}
\def\ip{{\rm Im}(\Phi)}
\def\tQ{{\tilde Q}}
\def\bi{{\vec{\beta}}^i}
\def\vp{{\left< \phi \right> }} 
\def\cm{{\cal M}}
\def\bZ{{\bf Z}}


\Title{hep-th/9601011, TAUP-2313-96}
{\vbox{\centerline{Exact Electric-Magnetic Duality in $N=2$}
\centerline{Supersymmetric QCD Theories}}}
\bigskip
\centerline{Ofer Aharony\foot{Work supported in part by the 
US-Israel Binational Science Foundation, by GIF -- the German-Israeli
Foundation for Scientific Research, and by the Israel Academy of 
Science.}$^,$\foot{Work supported in part by the Clore Scholars Programme.}
and Shimon Yankielowicz$^1$}
\vglue .5cm
\centerline{School of Physics and Astronomy}
\centerline{Beverly and Raymond Sackler Faculty of Exact Sciences}
\centerline{Tel--Aviv University}
\centerline{Ramat--Aviv, Tel--Aviv 69978, Israel}

\bigskip\bigskip

\noindent
We analyze the Coulomb phase of theories of $N=2$ SQCD with $SU(N_c)$ gauge
groups which are conjectured to have exact electric-magnetic duality. We 
discuss the duality transformation of the particle spectrum, emphasizing
the differences between the general case and the $SU(2)$ case. Some 
difficulties associated with the definition of the duality transformation 
for a general gauge group are discussed. We compute the classical monopole 
spectrum of these theories, and when it is possible we use it to check the 
consistency of the duality. Generally these theories may have phase 
transitions between strong and weak coupling, which prevent the 
semi-classical computation from being useful for checking the duality.

\Date{1/96}


\newsec{Introduction}

Electric-magnetic duality \om\ has played a major role in the exciting recent
developments in our understanding of gauge theories and string theories.
Generally, this duality is not an exact symmetry, but only a transformation
between different descriptions of
the low-energy (IR) theory. However, there are two classes of theories
which are conjectured to have an exact
electric-magnetic duality symmetry. These are the $N=4$ SYM theories
\osborn, and $N=2$ SQCD theories with a matter
content chosen such that the beta function vanishes perturbatively \hsw, and is
conjectured to vanish also non--perturbatively \refs{\swi,\swii}. 
Both of these theories are
also conjectured to be finite. The electric-magnetic duality in other $N=2$
SQCD theories may be derived by flowing from the scale invariant
theories \swii, and some sort of flow may also relate Seiberg's $N=1$ duality
\seiberg\ to the duality in scale invariant $N=2$ theories \ls.

For $N=4$ SYM theories,
the transformation of the $SL(2,\bZ)$ duality on the particle
spectrum for general
gauge groups is known since the late seventies \om. The electric charges in
this case sit in the root lattice of the gauge group, $\vec{e}=\sum_i
n_e^i \ai$, where the sum is over the simple roots of the gauge
group $G$. 
The magnetic charges sit in the root lattice of the dual gauge group,
$\vec{g} = \sum_i n_m^i {\ai\over{(\ai)^2}}$. The $SL(2,\bZ)$
duality acts on the (non-running) coupling constant
$\tau={\theta \over 2\pi} + {{4\pi i}\over g^2}$ of the high-energy
gauge group $G$, and is generated
by $T:\tau \to \tau+1$ and by $S: \tau \to -{1\over \tau}$. The generator
$S$ exchanges electric and magnetic charges, and exchanges
the roots with the dual roots : $\ai \leftrightarrow
{\ai \over (\ai)^2}$, up to a constant depending on the
normalization chosen for the roots. The gauge boson with electric charge
$\vec{\alpha}$ is thus transferred by $S$ to the magnetic monopole with
magnetic charge $\vec{\alpha} \over {\vec{\alpha}}^2$, which is exactly the
embedding of the 't~Hooft Polyakov monopole in the direction of the root
$\vec{\alpha}$ \bais.
This transformation exchanges the group $G$ with the
dual gauge group $G^{\vee}$. For $SU(N_c)$ the dual group is
$SU(N_c)/\bZ_{N_c}$, while for groups having roots of unequal length, the
dual group generally has a different algebra than the original algebra.
The verifications of this duality so far have involved the curve describing
the low-energy $U(1)^r$ gauge theory \refs{\swii,\dw}, 
and various partition functions \refs{\vw,\ggpz}, all
of which were found to be $SL(2,\bZ)$ invariant. Another verification of
the duality comes from checking that the BPS-saturated
particle spectrum of the theory is invariant under the duality.  
For $N=4$ SYM, the only BPS states which are only electrically charged are the 
gauge bosons, whose charge
$\vec{e}$ goes over all roots of the gauge group. The
tests made so far on the duality of the spectrum
have involved finding the $SL(2,\bZ)$ duals of these states
by a semi-classical analysis. As discussed below, the semi-classical
computation is sufficient for the $N=4$ SYM theory, 
since the spectrum is the same at strong and weak coupling.
For gauge group $SU(2)$ the BPS spectrum has been shown to be
consistent with the duality, for magnetic charge 2 by Sen \sen, and for
all magnetic charges by Porrati \porrati. For higher gauge groups it is
clear that the $S$-duals of the $W$ bosons exist semi-classically, 
since these are just the
embeddings of the 't~Hooft Polyakov monopole, but it has not yet been shown
that
the complete spectrum is indeed dual (for instance, 
that no other magnetic monopoles with zero electric charge exist).

$N=2$ SQCD theories with vanishing beta function are
known to be superconformally invariant and finite at the perturbative level
\hsw, and they are
conjectured (as we will assume from here on) to be superconformally
invariant and finite also non-perturbatively.
Seiberg and Witten have conjectured that these theories also have an exact
electric-magnetic duality for $SU(2)$ gauge group, and this has since been
generalized to higher gauge groups. However, so far the only
verification of the duality for higher gauge groups is in the
spectrum-generating curve of
the low energy $U(1)^r$ theory, which is invariant under a subgroup of
$SL(2,\bZ)$. Since massless
particles cause monodromies that may be read from the curve, 
the invariance of the curve implies that the particles which become
massless anywhere in the moduli space are also invariant under the
corresponding duality subgroup.
However, apriori it
says nothing about particles that are massive in all of moduli space, or
about particles that
are massless only together with other particles which are non-locally related
to them \refs{\ardo,\apsw}. In this case it is not obvious whether the quantum
numbers of the massless particles may be read from the curve.
In the $SU(2)$ case, the duality was also verified by checking
that the BPS-saturated particle spectrum is $SL(2,\bZ)$ 
invariant for all states which have
been computed semi-classically \refs{\swii,\suspect}. 
In this paper we wish to generalize
this to higher gauge groups. We will discuss only $SU(N_c)$ groups, but we
expect similar results to be true for all gauge groups.
 
Let us start by reviewing the situation for gauge group $SU(2)$
\refs{\swi,\swii}. In this case,
Seiberg and Witten conjectured, for $N_f=4$ (where $N_f$ is the number of
hypermultiplets in the fundamental representation), 
the existence of an $SL(2,\bZ)$
duality symmetry acting on $\tau={\theta \over \pi} + {{8\pi i}\over g^2}$
(note the factor of two difference from the previous case). This time there
are two kinds of states having only electric charge : the quarks with
charges $(n_m,n_e)=(0,\pm 1)$, in the vector representation of the $SO(8)$
flavor group, and the $W$ bosons with charges $(0,\pm 2)$ which are flavor
singlets. The $SL(2,\bZ)$ duality transformations were found to 
involve also outer automorphisms of the $SO(8)$ flavor group.
Since all non-singular points in the quantum moduli space are smoothly
connected to the semi-classical weak
coupling region, a semi-classical computation of the BPS spectrum is expected
to give the correct result throughout the moduli space.
The spectrum of states with $n_m=1,2$ was analyzed semi-classically by several
groups \suspect, 
and was found to be consistent with the duality. All states
which can be reached by a duality transformation on the elementary states
were indeed found, and, for these values of $n_m$, 
no other states were found. For general
$n_m$ the semi-classical analysis of the spectrum involves analyzing the zero
modes of a certain operator on the moduli space of $n_m$ monopoles, which
is a complicated hyperK\"ahler manifold. This analysis has not yet
been performed.

For general gauge groups much less is known about the $N=2$ duality. For all
simple non--exceptional gauge groups, the curve whose periods generate the
spectrum has been found by now for all interesting values of $N_f$
\refs{\ho,\aps,\as,\hanany}. The results are summarized in the introduction
of \as. The parameters of this curve in the scale invariant cases are the 
gauge coupling $\tau$ and gauge invariant polynomials built from the Higgs
field $\Phi$, all of which are uniquely defined only for weak coupling 
(different definitions may differ for strong coupling).
The gauge symmetry is
generically broken to $U(1)^r$ (where $r$ is the rank of the gauge group), 
and the gauge couplings are given by a matrix
$\tau_{ij}=\del_i \del_j \cal{F}$. At the classical level $\tau_{ij}$ is
just a constant matrix, proportional to the Cartan matrix of the group when
the $U(1)$ factors are defined to be aligned with the simple roots of the
gauge group. In the case of $SU(3)$ gauge group with $N_f=6$, on a 
non-singular subsurface of the moduli space, 
this is true also at the quantum level, as described below.
The low energy $U(1)^r$ gauge theory has an $Sp(2r,\bZ)$
transformation group, generated by $S$ transformations on each $U(1)$ factor
separately, by shifts of the various theta angles and by rotations
mixing the various $U(1)$ groups. A subgroup of this transformation group
keeps the matrix $\tau_{ij}$ proportional to the Cartan matrix (in the basis
mentioned above), and a subgroup
of this group may be an exact symmetry of the scale--invariant theory.
For gauge groups other than $SU(2)$, this subgroup turns out not to be
$SL(2,\bZ)$. This is revealed by looking at the curves and seeing 
which transformations of the
coupling constant leave them invariant. Instead, it 
is a subgroup of $SL(2,\bZ)$ which depends on the gauge
group. For $SU(N_c)$ ($N_c > 2$) and $SO(N_c)$ ($N_c > 4$), 
in the parametrization of the curves given in \as, the duality subgroup
acting on the coupling 
is generated by $S$ and by $T^2$. For
$Sp(N_c)$ it is generated by $T$ and by $ST^2S$. 
Note that $T^2$ is now the shift of the theta angle by $2\pi$, and that a 
shift of the theta angle by $\pi$ is not generally a symmetry.

In this paper we study the action of the duality transformations on the
particle 
spectrum of the $N=2$ SQCD theories, in an attempt to verify the existence
of an exact duality symmetry for all gauge groups. We begin in section 2
with a general description of the $N=2$ SQCD theories for general gauge
groups. In section 3 we check which $Sp(2r,\bZ)$ transformations preserve the
form of the classical coupling matrix $\tau_{ij}$. For $SU(3)$ gauge group
the classical coupling matrix is exact on a subsurface in moduli
space. In section 4 we compute
the classical monopole spectrum of the $N=2$ theory, finding the number of
bosonic zero modes around any classical monopole solution. In section 5 we
add also the fermionic zero modes, and describe the quantum numbers of the
semi-classical monopoles. In section 6 we check if the semi-classical
spectrum is consistent with electric-magnetic duality. Unfortunately, we
find that the semi-classical verification is only possible in a part of
moduli space in which we were not able to
complete the full semi-classical analysis (due
to mathematical difficulties). We show, however, 
that the semi-classical spectrum
could be consistent with the duality. In section 7 we analyze the theories
with non-zero bare quark masses, and see that also for these theories we
cannot verify or rule out the duality by a semi-classical analysis. We end
in section 8 with a summary and conclusions.

\newsec{General description of $N=2$ SQCD}

We consider $N=2$ supersymmetric gauge theories with
a gauge group $G=SU(N_c)$, 
and with $N_f$ hypermultiplets in the fundamental
representation (`quarks'), with $N_f$ chosen so that the beta function of the
theory vanishes perturbatively (for $SU(N_c)$ gauge groups $N_f=2N_c$).
The field content of these theories includes the $N=2$
vector multiplet, whose scalar component will be denoted by $\phi$. In terms
of $N=1$ superfields the $N=2$ vector multiplet consists of a vector
superfield $W_{\alpha}$ and a chiral superfield $\Phi$, both in the
adjoint representation of the gauge group. The
$N_f$ $N=2$ hypermultiplets consist of two chiral superfields,
$Q^i_a$ and $\tQ_i^a$, where $i=1,\cdots,N_f$ is a flavor index and
$a=1,\cdots,N_c$ is a color index. 
The superpotential in the $N=1$ language, for
zero quark masses (as we will assume until section 7), is
\eqn\spot{W = \sqrt{2} \tQ_i \Phi Q^i,}
suppressing the color indices. The flavor symmetry of this theory (for $N_c
> 2$) is $SU(N_f)\times U(1)_B$. The global
symmetry includes also the usual $SU(2)_R \times U(1)_R$ factors of the
classical $N=2$ theory. For $N_f=2N_c$, there is no perturbative 
anomaly in the $U(1)_R$ symmetry, and we will assume that it remains 
unbroken in the full quantum theory (except for spontaneous breaking).

We will be interested here only in the Coulomb phase of these theories, in
which the only field obtaining a VEV is $\phi$. The equations of motion
imply $[\phi,\phi^{\dagger}]=0$, so that in a vacuum we can always diagonalize
the matrix $\vp$, i.e. choose it to be in the Cartan subalgebra of the gauge
group. Choosing the basis of the Cartan subalgebra to be the simple
roots of the gauge group, we can thus take
\eqn\phivev{\vp = \sum_{i=1}^r a_i \ai}
where $r$ is the rank of the gauge group. Classically,
the moduli space is labeled by
the value of the $a_i$ up to gauge transformations. After choosing
$\vp$ in the Cartan subalgebra, we are left only with the freedom to perform
Weyl transformations. 

For $SU(N_c)$ gauge groups the Coulomb phase moduli space of vacua may be 
parametrized by the gauge invariant operators $u_k = {1\over
k}\vev{\Tr(\Phi^k)}$
for $k=2,\cdots,N_c$. 
Equivalently, one may use $s_k$, the symmetric polynomials in the eigenvalues
of $\Phi$. Classically,
these determine the $a_i$ up to Weyl transformations.
In the theories we analyze we expect to have exact scale invariance and
exact $U(1)_R$ invariance. Hence, the particle spectrum at the point 
$\{ u_k \}$ in
the moduli space should be the same as the particle spectrum at the point $\{
\lambda^k u_k \}$ for any complex $\lambda$. Dimensionless parameters, such
as the effective gauge coupling, may still depend (for $N_c > 2$) 
on dimensionless ratios
of the $u_k$, such as $u_3^2/u_2^3$. Unlike the case of gauge group
$SU(2)$, for higher rank groups the $U(1)_R$ symmetry does not enable us to
choose $\vp$ to be real. At best, we can (and will) choose $\vp \cdot
\vec{\alpha}$ to be real for some particular $\vec{\alpha}$ in the Cartan
subalgebra.

At a generic point in moduli space, the VEV of $\phi$
breaks the gauge group to $U(1)^r$, and a low energy effective theory may
be written in terms of $U(1)$ vector multiplets 
$(A_i,W_i)$ ($i=1,\cdots,r$), which are the only massless fields. We can
choose the basis of the $U(1)^r$ gauge group such that the VEV of the
scalar component of $A_i$ is exactly $a_i$ defined above. The $N=2$
effective lagrangian takes the form
\eqn\leff{L_{eff}=Im{{1\over{4\pi}}} \left[ \int d^4 \theta \del_i {\cal
F}(A) {\bar A}^i + \half \int d^2 \theta \del_i \del_j {\cal F}(A) W^i W^j
\right]}
where $\cal F$ is a holomorphic prepotential \seibergp. 
Classically, in the basis we chose, $\cal F$ is
proportional to
\eqn\fcl{{\cal F}_{cl}(A) \propto \tau \sum_{i,j} A_i A_j C_0^{ij}}
where $\tau = \theta / \pi + 8\pi i / g^2$ is the bare gauge coupling
(which is well defined for $N_f=2N_c$) and $C_0$ is the
Cartan matrix of the gauge group ($C_0^{ij} = {{2\vec{\alpha}^i \cdot
\vec{\alpha}^j} \over {(\vec{\alpha}^j)^2}}$). We will choose to normalize
the coupling so that $\theta$ is the coefficient of the topological term in
the lagrangian\foot{This convention differs by a factor of $N_c$ from the
convention used when describing the large $N_c$ limit of these theories.}.

When $u_i=0$ for $i=2,\cdots,N_c-1$, and only $u_{N_c}$ is non-zero,
the one-loop correction to the low energy coupling $\tau_{ij}=\del_i \del_j
{\cal F}$ is a constant matrix\foot{This point has been stressed in a
recent paper \newmn.}, 
essentially because there is only one scale in the theory, and the masses
of all particles are multiples of this scale. For $G=SU(3)$ the constant
one-loop correction vanishes, and the classical result is perturbatively
exact on this subsurface. Other surfaces with only
one scale also exist, but on all of them, except
this one, we have some massless quarks, and then the effective
coupling runs below the scale of the massive particles. We will denote
this subsurface of the moduli space by $\cm$.
In the $N=4$
theory, the one-loop correction vanishes throughout moduli space, due to the
cancelation between the $N=2$ vector multiplet and hypermultiplet, 
but this is not generally true in the $N=2$ theory. The higher 
perturbative corrections always vanish. For $SU(2)$ gauge group
there are no
non-perturbative corrections to this expression in the scale invariant
theories. It is reasonable to expect that this will be true also for
larger gauge groups. 
For $G=SU(3)$ on the subsurface $\cm$,
the expression for the dual variables $a_D^i$ is then exactly given by
\eqn\ad{a_D^i = \del_i {\cal F} = K \tau C_0^{ij} a_j}
where $K$ is a constant.
For the $N=4$ theories, \fcl\ and \ad\ are exact 
throughout the whole moduli space.

The states of this theory generally carry electric and magnetic charges.
The electric charges live in the weight lattice of the gauge group.
With an appropriate normalization, we can write
${\vec e} = \sum_{i=1}^r n_e^i \vec{\mu}_i$, where the weights 
$\vec{\mu}_i$ are defined by 
$\vec{\mu}_i \cdot \vec{\alpha}^j = \half
\delta_i^j (\vec{\alpha}^j)^2$ and the charges $n_e^i$ are all integers. 
The magnetic charges ${\vec g}$
(appropriately normalized) must then
satisfy the Dirac quantization condition
${\vec e} \cdot {\vec g} = 2\pi n$ for all ${\vec e}$ of purely
electric states. This implies ${\vec g} = 4\pi \sum_{i=1}^r n_m^i \bi$,
where $\bi = {\ai \over {(\ai)^2}}$ and the $n_m^i$ are all integers. Thus,
the magnetic charges lie in the root lattice of the dual gauge group
(we will normalize $\vec{g}$ without the $4\pi$ factor from here on to
simplify the notations). 
Note that this quantization is different than the case of an unbroken
simple gauge group, when the magnetic charges may lie in the weight lattice of
the dual gauge group. The
mass of a state is bounded by the BPS bound, which for zero quark masses is
given by
\eqn\bpsbound{M \geq \sqrt{2} |Z| = \sqrt{2} |n_e^i a_i + n_m^i a_D^i|,}
with equality only for states in small representations of the $N=2$ algebra.

The low energy action has an $Sp(2r,\bZ)$ group of
transformations, acting on $(a,a_D)$ (now viewed as vectors of $r$
elements) by
\eqn\sptrans{\pmatrix{a \cr a_D \cr} \to \pmatrix{A & B \cr C & D \cr}
\pmatrix{a \cr a_D \cr}}
and on $(n_e,n_m)$ in an appropriate way so that $Z$ of \bpsbound\  is
invariant. The matrix of gauge couplings $\tau_{ij} = \del_i \del_j {\cal F}$
is transformed by \sptrans\ as $\tau \to (C + D \tau) (A + B \tau)^{-1}$.
In general these transformations are not expected to be exact
symmetries of the $N=2$ SQCD theory, 
but some subgroup of $Sp(2r,\bZ)$ may in fact
be an exact symmetry, as discussed in the next section. Such a symmetry
should preserve the matrix form of $\tau_{ij}$, changing only the
gauge coupling coefficient.

An important difference which arises when the rank of the gauge group is
larger than one, is that even in the scale invariant theories we have
singular surfaces of real codimension one on which states of the theory are
only marginally stable. These occur when the central charges $Z$
corresponding to two different states of the theory have the same complex
phase. In addition to these surfaces, there exist singular surfaces of real
codimension two, corresponding to massless fields, which can be either
massless vectors (which occur for instance
when the gauge symmetry is not completely broken to an
abelian subgroup) or massless hypermultiplets. The codimension two
singular surfaces generate $Sp(2r,\bZ)$ monodromies when going around them.
On the other hand, the spectrum of the theory can jump upon crossing
codimension one
singular surfaces on which states are only marginally stable \jumps. 
States which are only marginally stable on the
singular surface may exist as stable particles on one side of the surface
but not on the other.
This happens for instance in the $SU(2)$ case, even for $N_f=0$ \swi.
As we will see in sections 4 and 5, 
the semi-classical computation of the spectrum
indicates that such jumps do indeed occur even in the
semi-classical region.

In the $SU(2)$ theory with $N_f=4$ \swii, we do not cross any singular
surfaces of this type
when the coupling is changed from strong coupling to weak coupling,
because the expression $a_D = \tau a$ is exact. In the $N=2$ theories of
$SU(N_c)$ for $N_c > 2$, due to quantum corrections, this
is no longer the case at a generic point in moduli space. Hence, the
particle spectrum at weak coupling, which we can compute by a
semi-classical analysis, is not necessarily the same as the spectrum at
strong coupling. From the known form of the low-energy curve given
below, we can see that there are no codimension one singular surfaces 
which intersect the subsurface
$\cm$. However, for $N_c > 3$ we cannot show that there are no codimension 
two singular surfaces which intersect $\cm$, so we cannot be sure that
the spectrum does not change when we go from weak to strong coupling.
Only for $SU(3)$, the relation \ad\ is exact along the
surface $\cm$. Therefore, along this surface, the 
spectrum should be the same
at strong and weak coupling, and its semi-classical computation
should be self-dual.
We will only be able to perform semi-classical tests
of the duality near this special surface in moduli space.
In principle, the exact form of the singular
surfaces may be computed from the low-energy curves, and then we can
tell exactly where the semi-classical calculation is reliable and where it
is not. Since we have not done this, we will have to restrict ourselves
to rigorously checking the duality only
for $SU(3)$ near the surface $\cm$, where we know that
the spectrum is the same at strong and weak coupling. For $N_c > 3$ we
cannot prove that there is a region of moduli space in which 
no phase transitions occur. However, we will see below 
that the spectrum along $\cm$ is consistent with the
duality. Thus, it is possible that 
no phase transitions occur along this surface.
Note that for the
$N=4$ theories, there are no quantum corrections and
\ad\ is valid throughout the moduli space. Therefore, in these
theories the spectrum is the same at strong and weak coupling for all gauge
groups, and the semi-classical spectrum should be self-dual. 

\newsec{Duality transformations of $N=2$ SQCD}

Several groups have constructed curves describing the
low-energy physics in the
Coulomb phase of $N=2$ SQCD theories for general gauge groups. For
$SU(N_c)$ gauge groups with $N_f=2N_c$, the curve (in the parametrization of
\aps) is given by
\eqn\curve{y^2 = \vev{\det(x-\Phi)}^2 + 
4h(\tau)(h(\tau)+1) \prod_{j=1}^{2N_c} (x
- m_j - {1\over N_c} h(\tau) \sum_{k=1}^{2N_c} m_k),}
where the function $h(\tau)$ is given by $h(\tau) = {{\theta_2^4(\tau)}
\over {\theta_4^4(\tau) - \theta_2^4(\tau)}}$ (for weak coupling $h(\tau)
\sim 16 e^{i\pi \tau}$), and the $m_k$ are the masses of the quarks.
The parameter $\tau$ appearing in the curve is related to the (non-running) 
gauge coupling of the high-energy non-abelian gauge group,
and should be equal to it at least for weak coupling. 
This curve has two symmetry transformations which leave it invariant.
One is the transformation $T^2:\tau \to \tau + 2$, corresponding to
$\theta \to \theta + 2\pi$ which is obviously a symmetry of the theory. 
Another symmetry transformation is
$S:\tau \to -1/\tau$, which is assumed to be related to electric-magnetic
duality. The $S$ transformation 
also inverts the sign of the singlet quark mass but not
of the $SU(N_f)$-adjoint masses, corresponding to an outer automorphism of
the flavor group which inverts the $U(1)_B$ charge and does not act on
$SU(N_f)$. 
Both of these symmetry transformations do not act on the gauge invariant
variables $u_k$ (appearing in the determinant in \curve), 
which are the other parameters of the curve.

The relation between the parameter $\tau$ of the curve and the actual gauge
coupling is complicated, and is not a one-to-one transformation. This can
easily be seen by comparing the parametrization \curve\ with other
parametrizations of the curve. For $SU(3)$ another parametrization was given
in \mn, which had duality transformations $S$ and $T$ which satisfied
$(ST)^6=I$, while no such relation is satisfied by the symmetry generators
of \curve\foot{Another way to explain this would be if the $S$ transformation
of \mn\ is not easily visible in the parametrization \curve\ and vice
versa.}. For $SU(2)$, we can compare the $S$ transformation above to the
transformations given by Seiberg and Witten \swii, and see that $S$ above
corresponds in their parametrization to a transformation of the form $TST$,
which does not square to unity when acting on the coupling defined in \swii.
The $SU(2)$ example also clearly shows that the transformations which are 
easily read
from \curve\ are not necessarily the most general duality transformations.
For instance, the $S$ transformation of \swii\ is not easily visible in the
parametrization \curve\ of the curve.
We will, therefore, look in this paper for more general possible forms of
duality transformations. In general, these dualities could also involve
different outer automorphisms of the $SU(2N_c)\times U(1)_B$ flavor group.
The $SU(2N_c)$ group has just one non-trivial outer automorphism, which
conjugates the $SU(2N_c)$ representations, but we could also have
inversions of the baryon number charge, and shifts of the baryon number
charge by the $\bZ_{2N_c}$ charge corresponding to the
center of the $SU(2N_c)$ group.
In general, the duality transformations
may also act on the $u_k$, but we will look only
for transformations which leave the $u_k$ invariant.

We consider now general
electric-magnetic duality transformations, which transform the $N=2$ SQCD
theory with high energy coupling $\tau$ to the same theory with high energy
coupling $\tau'=f(\tau)$.
The respective low energy theories should be
related by an $Sp(2r,\bZ)$ transformation.
We begin by assuming that the classical relation \ad\ holds both before and
after the duality transformation. As discussed above, this is true also
quantum mechanically for the cases of $G=SU(2)$, $G=SU(3)$ along the
surface $\cm$ \foot{Since we assume that the
duality transformation does not change the $u_k$, it leaves the surface
$\cm$ invariant.} 
and $N=4$ SYM. 
For the other cases the quantum corrections are important and will be
discussed at the end of this section.

With this assumption, we should look for $Sp(2r,\bZ)$ transformations
which leave the relation \ad\ invariant, up to a possible change of
$\tau$. Equivalently, they should preserve
the matrix form of the coupling matrix $\tau_{ij} = K \tau C_0^{ij}$.
Let us now take a general
$Sp(2r,\bZ)$ transformation, and a general transformation $\tau \to f(\tau)$,
and check which transformations leave the relation \ad\ invariant. We will
work in matrix notation, as in equation \sptrans. After the transformation,
$a_D$ is given by $Ca+Da_D$ and $a$ is given by $Aa+Ba_D$. Equation
\ad\ then becomes
\eqn\first{Ca+Da_D = K f(\tau) C_0 (Aa+Ba_D).}
Plugging in the relation \ad\ for the original variables, we find the matrix
equation
\eqn\second{C+K\tau D C_0 = K f(\tau) C_0 (A + K \tau B C_0)}
which must be satisfied for all $\tau$. 

Since $f(\tau)$ should be a holomorphic function, we can take derivatives
of \second\ with respect to $\tau$. By demanding equality for all $\tau$ we
then find that all matrices appearing in \second\ ($C,DC_0,C_0A$ and $C_0BC_0$)
must be proportional to each other, and $f(\tau)$ must be of the form
$f(\tau) = (a \tau + b) / (c \tau + d)$. The fact that all these matrices
are proportional, means that 
if we start from a state whose magnetic and electric charges (the vectors
$\vec{g}$ and $\vec{e}$ defined above) are proportional to each other, 
we will remain with proportional electric and magnetic charges. 
This results from the fact that $C_0$ is proportional,
for simply laced groups, to the matrix 
transforming the electric charge basis to the magnetic charge basis. 
In particular,
states which are charged only electrically or only magnetically are necessarily
transformed into states whose electric charge is in the same direction as
their magnetic charge.

Let us analyze first the case in which $B$ is zero. Then
equation \second\ becomes
\eqn\bzero{C+K\tau D C_0 = K f(\tau) C_0 A.}
Clearly, the only possible solutions to this equation for all $\tau$ are of
the form $f(\tau) = a_0 \tau + b_0$. Since $A^T D=I$ in this case (where
$I$ is the identity matrix), we can easily show that $a_0=\pm 1$, and
obviously only $a_0=1$ is physically relevant, since $b_0$ is real and the
imaginary part of $\tau$ is always non-negative. These transformations are
exactly the transformations of the form $T^{2n}$ described above, and we see
that they can indeed leave the theory invariant. Since $T^2 : \tau \to \tau
+ 2$ is supposed to take $\vec{e} \to \vec{e} + \vec{g}$ (when the length
of the roots is
normalized to one), we find that $K = 1/2$ for all $N_c$ in our conventions.
Obviously these transformations do
not exchange electric and magnetic charges; for electric-magnetic duality
we must obviously have non-zero $B$ which is the next case we shall
analyze.

Since we found that all matrices in \second\ are proportional, we can write
(for non-zero $B$) $A = q_1 B C_0$, $C = q_3 C_0 B C_0$, and $D = q_4 C_0
B$, where $q_1, q_3$ and $q_4$ are constants, which must obviously be
rational numbers, since all matrices are integer valued. 
The transformation on $\tau$ is then
\eqn\trans{\tau \to f(\tau) = {{q_3 + K q_4 \tau} \over {K(q_1 + K\tau)}}.}
Using the $Sp(2r,\bZ)$ relation $A^T D - C^T B = I$ we can see that 
\eqn\cbcb{(q_1 q_4 - q_3) C_0 B^T C_0 B = I.}
This equation essentially means that $B$ should preserve the form of the
lattice of charges as a sublattice of the Cartan subalgebra.
Taking the determinant of this equation we find that
\eqn\problem{(q_1 q_4 - q_3)^r = 1 / (\det(C_0) \det(B))^2.} 

By analogy with the
$SU(2)$ case, we will first look for solutions in which $A=D=0$,
i.e. there is no mixing between electric and magnetic charges. 
In this case, taking the determinant of $C = q_3 C_0 B C_0$, we find that
$\det(C) = (-1)^r / \det(B)$, but both determinants must be integers, hence
$\det(B) = \pm
1$. Plugging this into \problem, we find that there are no rational
solutions for $q_3$ for $SU(N_c)$ groups with $N_c > 3$ (recall that
$\det(C_0) = N_c$). Thus, for these groups, any duality transformation
must mix electric and magnetic charges in a more
complicated way than for $SU(2)$. For generic $N_c$ the smallest value of
$\det(B)$ for which \problem\ has a rational solution is
$\det(B)=\pm N_c^{N_c-2}$, and then 
$q_1 q_4 - q_3 = 1 / N_c^2$.
We can always choose, for instance, $B=N_c C_0^{-1}$ (which is an integer
matrix) and satisfy all the equations. For particular values of $N_c$
smaller values of $\det(B)$ are also possible. All solutions
of \cbcb\ give rise to transformations which preserve the low energy
effective action and the form of the low energy couplings.
To find which of the solutions are exact symmetries
of the theory we must go beyond the low energy action, for instance by
studying the particle spectrum of the theories, which we do in the next 
sections.

The duality transformation of the charges is
the inverse of the transformation on $(a,a_D)$. For the
transformation matrices given above we find that
\eqn\transn{\pmatrix{n_e \cr n_m \cr} \to 
\pmatrix{q_4 C_0 B & -q_3 C_0 B C_0 \cr -B & q_1 B C_0 \cr }
\pmatrix{n_e \cr n_m \cr}.} 
Denoting $\vec{g}_i = -B_{ji} \vec{\beta}_j$, and normalizing all root
lengths to
unity (which is possible for simply laced groups), we find that the
transformation of the basis of the charge lattice is given by
\eqn\atrans{\eqalign{ \vec{e} = \vec{\mu}_i & \to 
\vec{e} = -q_4 \vec{g}_i \qquad \quad \vec{g} = \vec{g}_i \cr
\vec{g} = \vec{\beta}_i & \to \vec{e} = q_3 (C_0)_{ji} \vec{g}_j \quad
\vec{g} = -q_1 (C_0)_{ji} \vec{g}_j. \cr }}
Using \cbcb\ it is easy to verify that this transformation preserves the
symplectic product $\vec{e}_1 \cdot \vec{g}_2 - \vec{g}_1 \cdot \vec{e}_2$
between any pair of vectors. The freedom to perform $T^2:\tau \to \tau+2$
transformations corresponds here to a freedom to transform the rational
numbers $q_i$ by 
$q_3 \to
q_3+2Kq_4; q_1 \to q_1+2K; q_4 \to q_4$ or by $q_3 \to q_3+2Kq_1; q_4 \to
q_4+2K; q_1 \to q_1$. Both transformations of course preserve $(q_1
q_4-q_3)$. Thus, up to $T^2$ transformations, we can always choose the
absolute values of $q_1$ and $q_4$ to be no larger than $\half$. 

Up to now we did not add any constraints on the duality from the flavor
quantum numbers of the various states. As we discuss in section 6, for all
states in the theory the charge ($n$-ality) under the center of the
$SU(2N_c)$ flavor group is equal (modulo $N_c$) to the charge under the
center of the $SU(N_c)$ gauge group (given by $\sum_i i n_e^i$). 
For $N_c > 2$, since there is only one
$SU(2N_c)$ representation (up to conjugation) of size $2N_c$, and the
duality preserves the number of states, it seems that the duality must also
preserve the $SU(N_c)$ $n$-ality of states, or invert it if an $SU(2N_c)$
outer automorphism is also involved. From \atrans\ we can easily see that
for this to happen, $q_4$ cannot be an integer, since then all weights would
transform to roots, whose $n$-ality is zero. In fact $q_4$ must be of the
form $p/q$ where $q$ is an integer multiple of $N_c$, and this will constrain
the possible transformations. We assumed here that the monopoles carry no
charge under the center of the gauge group -- if this is not correct then
we should add also the $n$-ality of the monopoles to this discussion.

The transformation matrix $B = N_c C_0^{-1}$ seems to be the most natural
choice. In this case, the transformation is similar to the $N=4$ 
transformation,
taking electric charge vectors to magnetic charge vectors which are
proportional to them (as vectors in the Cartan subalgebra). 
For this transformation we find
$q_1 q_4 - q_3 = 1 / N_c^2$, to which the simplest solution giving an
$Sp(2r,\bZ)$ transformation which preserves or inverts
the $n$-ality of electric charge vectors is obviously 
$q_1 = q_4 = \pm 1/N_c, q_3=0$.
Equation \atrans\ in this case becomes
\eqn\btrans{\eqalign{ \vec{e} = \vec{\mu}_i & \to 
\vec{e} = \mp \vec{\mu}_i \qquad \vec{g} = N_c \vec{\mu}_i \cr
\vec{g} = \vec{\beta}_i & \to \vec{e} = 0 \qquad \quad
\vec{g} = \mp \vec{\beta}_i. \cr }}
This looks just like a conjugation by $S$ of a $T^{N_c}$ transformation.
As we shall see below, this transformation agrees
with the semi-classical monopole spectrum of the theory in the parts of moduli
space for which the semi-classical spectrum should be self-dual. However,
this is also true for many other transformations, so it is not clear that
this is indeed the ``correct'' duality transformation. We might try to
check if this duality transformation is consistent with the flow from
$SU(N_c)$ gauge group to $SU(N_c-1)$ gauge group, as described in \aps.
If we naively compare the $a$'s and $a_D$'s of the two theories, we find
that in fact this transformation is not consistent with the flow (we find
for the $SU(N_c-1)$ theory $B = N_c {\tilde C}_0^{-1}$ instead of 
$B = (N_c - 1)
{\tilde C}_0^{-1}$). However, since the duality takes us to a strong coupling
theory, the quantum corrections should be important, and
we cannot trust this computation. For instance, for $G=SU(3)$,
the flow takes place far from the surface $\cm$ in moduli space
where \ad\ is correct quantum mechanically.
Hence, we cannot
really constrain the duality transformations by using this flow. 

For $N_c=3$, another possible duality transformation is $q_1=q_4=0$ and
$B=\pmatrix{1 & 0 \cr 1 &
1 \cr}$ up to Weyl transformations. This gives the duality
transformation described in \mn, 
taking $\tau$ to $-4/(3\tau)$. However, this transformation does not
preserve the $n$-ality of charge vectors, and in any case
for higher gauge groups
we do not have analogs of this transformation. Therefore, it seems that $B =
N_c C_0^{-1}$ is a more reasonable guess.

While it is possible to find consistent duality
transformations for all $N_c$, it
is not generally possible to find transformations whose square equals the
identity (acting on the coupling constant $\tau$), 
as seems to be implied from the
transformations of the curves. This can easily be seen, for instance, for 
$N_c=6$, which
is one of the cases in which we can see from \problem\ that $\det(B)$
must be a multiple of $N_c^{N_c-2}$
(for odd $N_c$ we can always find smaller solutions of 
\problem). In fact we can always write in this case $\det(B)=n^{N_c-1}
N_c^{N_c-2}$ for an integer $n$.
For the transformation squared to give unity (acting on $\tau$), $q_4$ must
equal $(-q_1)$, and then equation \problem\ becomes
$(-q_1^2 - q_3)^{N_c-1} = 1 / (n
N_c)^{2(N_c-1)}$, and, therefore, 
$q_1^2 + q_3 = -1 / (n N_c)^2$. Now, we can compute
the determinant of $C$. Using this relation it turns out to be 
$\det(C) = (-{1\over{(nN_c)^2}} - q_1^2)^{N_c-1} N_c^{N_c}$, which must be an 
integer.  Since $N_c$ is not a rational number to the $(N_c-1)$'th power for
such $N_c$, this integer must necessarily be a multiple of $N_c$, leading
to $({1\over{(nN_c)^2}}+q_1^2)N_c$ being an integer. Defining $r = n N_c
q_1$, which must still of course be a rational number, we find that $r^2+1$
must be an integer multiple of $n^2 N_c$. For $N_c=6$, taking this equation
modulo $6$, we find a contradiction, since there are no rational solutions
modulo $6$ to $r^2+1=0$. Thus, in this case there is no legal
transformation which squares to unity. Note that if we use $B = N_c
C_0^{-1}$, then for odd $N_c$ we can choose $q_4 = -q_1 = \pm 1 / N_c$ which
gives a transformation that does
square to unity, but this is not possible for even values of $N_c$.

Let us now briefly comment on the differences between this case and the
$N=4$ case,
for which electric-magnetic duality is believed to work for all gauge
groups, taking the gauge group to its dual by $\ai \leftrightarrow 
\bi$. In this case,
the electric charges are all in the root lattice of the gauge group, so
that the definition of the $Sp(2r,\bZ)$ transformations above is changed. The
matrix $C$ appearing in the transformation of the magnetic to the electric
charge numbers must now be a multiple of the Cartan matrix $C_0$,
which transforms the weights to the roots. The matrix $B$ is allowed to be 
non-integer acting on
the weights but must still be an integer when acting on the roots. Thus, in
this case we can write $C = C_0 \tilde{C}$ and $B C_0 = \tilde{B}$ where
$\tilde{B},\tilde{C}$ are integer valued matrices. Plugging this into
equation \second\ for $A=D=0$
we find that necessarily $\tau \to b_0/\tau$ and
$\tilde{C} = K^2 b_0 \tilde{B}$. In this case, however, we find
$\det(\tilde{C}) \det(\tilde{B}) = (-1)^r$, so that $K^2 b_0 = -1$ and
there are no problems in finding appropriate duality transformations. In
particular, $\tilde{C}=-\tilde{B}=I$ and $A=D=0$
gives the usual $S$ duality transformation of the $N=4$ theory, taking
$\tau \to -4 / \tau$ in our normalization of the coupling. 
Of course, in the $N=4$ case there are no $n$-ality requirements analogous
to those we discussed above, since all states have zero $n$-ality.
We run
into problems in the $N=2$ case because in this case the electric charge
lattice (which is the weight lattice of the gauge group) is generally not
isomorphic to the magnetic charge lattice (which is the root lattice of the
dual gauge group).

Our discussion so far assumed the validity of the classical relation
\ad. For the $N=4$ and $G=SU(2)$ cases, this relation is valid quantum
mechanically, hence our analysis is exact. In the $G=SU(3)$ case, the relation
\ad\ is valid quantum mechanically along the surface $\cm$, which is
non-singular, and our analysis is still valid there. We have
assumed that the
surface $\cm$ is invariant under duality transformations, which is
reasonable since it is the only surface along which the ${\bf S}_3$ Weyl
symmetry is broken to $\bZ_3$. For higher gauge groups there are no regions
in the quantum moduli space for which \ad\ holds\foot{As pointed out in a
recent paper \newmn.}, and our analysis
is not valid in the quantum theory. Since the duality is a quantum effect,
it is not clear that the classical part of the prepotential should be
invariant by itself. However, since the transformations we are dealing with
form a discrete group, it seems reasonable to expect that this will hold.
In this case, the constraints that we have derived above will still be
valid in the full quantum theory.

For $SU(N_c)$ gauge groups with $N_c > 3$ along the surface $\cm$, 
there is a constant one-loop
correction to the coupling matrix $\tau_{ij}$.
We can try to include this correction and repeat the analysis performed
above. However, since this matrix consists of irrational
numbers, we find that including this correction leaves no possible duality
transformations, except for the trivial $\tau \to \tau + 2$ transformation.
There are several possible ways to interpret this result. First, it is
possible that for these gauge groups, along the surface $\cm$, there are
also non-perturbative corrections to the prepotential, and these must also
be included. Second, it is possible that for $N_c > 3$ the duality does   
not preserve the surface $\cm$, though it seems that this is the only
surface along which the Weyl symmetry breaks to a $\bZ_n$ subgroup.
Finally, it is possible that there is no exact $S$ duality for these gauge
groups. The first possibility seems to be the most reasonable one.
For the scale invariant cases there are no rigorous arguments against the
appearance of non-perturbative corrections, as long as these do not break
the scale invariance. In the $SU(2)$ and $SU(3)$
cases, such corrections must have the same matrix form as the classical
coupling matrix, since it is the only matrix invariant under the $\bZ_n$
Weyl transformations. Therefore, we can always swallow them in the
definition of the coupling constant for strong coupling. For $N_c > 3$
other matrices may appear, and we need to understand better the
non-perturbative corrections in order to perform the full quantum analysis
of the duality.

\newsec{The classical monopole spectrum}

Next, we would like to compute the classical monopole spectrum of the
$N=2$ SQCD theories.
We will show that classically there is an equivalence between
BPS monopole solutions for real and complex Higgs fields, so that we may
use old results \refs{\weinberg,\weinbergg} 
on the classical moduli space for real Higgs
fields. By this we mean that for
every BPS monopole solution for real Higgs fields there exists a solution
for complex Higgs fields (with any VEV for the imaginary part of the Higgs
field in the Cartan subalgebra), and for every solution for complex
Higgs fields there is a solution for real Higgs fields. The
dimension of the monopole moduli space, i.e. the number of bosonic zero
modes, is, therefore, the same in both cases\foot{The exact relation 
between the
two cases involves a Weyl transformation as described later in this section.}.

Let us first derive the BPS bound for complex Higgs fields, and see what
the BPS equations are in this case. The analysis will be purely classical
throughout this section, and we will concentrate on solutions with no
electric charge. We will take the action to be the bosonic
part of the $N=2$ SYM lagrangian,
\eqn\action{S = \int d^4x \Tr \{ 
-{1\over 4}(F_{\mu \nu})^2 + {1\over 2}(D_{\mu} 
\rp)^2 + {1\over 2}(D_{\mu} \ip)^2 - {1\over 2}([\rp,\ip])^2 \}. }
In \action\ we have chosen, for simplicity, the theta angle to be zero,
and we normalized the gauge fields so that the gauge coupling is
one. The gluinos and the quark superfields are not expected to affect the
monopole solutions, except via zero modes upon quantization. The energy
density derived from this action is
\eqn\energy{\eqalign{U = \half \int d^3 x \Tr 
[ & (E_i)^2 + (B_i)^2 + (D_0 \rp)^2 +
(D_i \rp)^2 + (D_0 \ip)^2 + \cr & (D_i \ip)^2 + ([\rp, \ip])^2 ] = \cr =
\half \int d^3 x \Tr [ & (B_i - D_i \rp + i D_i \ip) (B_i - D_i \rp - i D_i
\ip) + \cr & 2 (B_i D_i \rp) + ([\rp, \ip])^2 + (E_i)^2 + (D_0 \rp)^2 + (D_0
\ip)^2 ].  \cr
}}

All components of the energy density written above are clearly
non--negative, except
for $\int d^3 x \Tr(B_i D_i \rp)$, which is exactly $v$ times 
the real part of the magnetic
charge, $Q_M = v^{-1} \int d^3 x \del_i(\Tr(B_i \Phi))$, where $v$ is the
absolute value of the asymptotic Higgs field. Therefore, we find that $U \geq
v {\rm Re}(Q_M)$, with equality only when all other parts of the energy
density vanish. However, there is a chiral symmetry freedom which enables us to
choose the phase of $\Phi$ (classically and also quantum
mechanically when the beta function vanishes), and we derive the best bound
by choosing the phase to make $Q_M$ positive and real. Thus, generally we
find that
$U \geq v |Q_M|$. This is the BPS inequality for states with no electric
charge, and we are looking for solutions to the classical equations of
motion for which the inequality is saturated, i.e. $U = v |Q_M|$. From
equation \energy, we deduce that if we choose the phase of the Higgs field as
defined above, such solutions must satisfy (locally)
\eqn\bps{\eqalign{
B_i & = D_i \rp \cr
0 &= D_i \ip \cr
0 &= [\rp, \ip] \cr
E_i &= D_0 \rp = D_0 \ip = 0. \cr}}
These equations generalize the relations $B_i=D_i \phi, E_i = D_0 \phi = 0$
which exist for a real Higgs field $\phi$. Obviously, any solution (in the
gauge $A_0^a = 0$) will be time independent, and we will assume this from
here on. Relaxing the assumption of 
zero electric charge we would find that both
$B_i$ and $E_i$ are proportional to $D_i \rp$. Thus, the electric and
magnetic charge vectors of classical BPS-saturated
dyons are always proportional to
each other.

Our goal in this section
is to relate the solutions to \bps\ which exist for real and
complex Higgs fields. We will look for local solutions, for which the
asymptotic magnetic field decays as $1/r^2$. The solutions
are then characterized by the asymptotic values of $\Phi$ and
of the magnetic field along (for instance) the $z$ axis. 
These asymptotic values commute,
since $D_i \Phi$ vanishes asymptotically. Therefore, both of them may be
chosen to be in the Cartan subalgebra. Following E. Weinberg \weinberg\
we will denote
the asymptotic value of the Higgs field along the $z$ axis by $\Phi_0 = v
{\vec{h}}\cdot{\vec{H}}$, where ${\vec{H}}_i$ is a basis for the Cartan
subalgebra, and $\vec{h}$ is a vector of unit absolute value. The asymptotic
behavior of the magnetic field along the $z$ axis will be of the form 
$B_z = {{\vec{g}} \cdot {\vec{H}}} / z^2$, 
where $\vec{g}$ satisfies the
quantization condition ${\vec{g}} = \sum_i n_m^i {\ai \over
{(\ai)^2}}$.  The vectors ${\vec{g}}$ and ${\vec{h}}$ are defined by this
procedure up to Weyl transformations, and the magnetic charge $Q_M$ (defined
above) is given by $Q_M = 4\pi \vec{g} \cdot \vec{h}$. 
In Weinberg's analysis \weinberg\ $\vec{h}$ is real, and
the Weyl transformation freedom is
used to set ${\vec{h}}\cdot \ai \geq 0$ for all simple roots $\ai$.
This removes any ambiguities for the case in which only an abelian symmetry
remains unbroken, i.e. $\vec{h} \cdot \vec{\alpha} \neq 0$ for all roots
$\vec{\alpha}$.

First, let us show that any classical solution for complex Higgs fields
gives a solution for real Higgs fields with the same magnetic charge.
This is obvious, since, as described above, we can always use a $U(1)_R$
transformation to choose the phase of the Higgs field 
in such a way that equation \bps\ is satisfied. This is achieved by
choosing
the phase of the vector $\vec{h}$ above so that ${\vec{g}} \cdot {\vec{
h}}$ is real and positive. Then,
simply by setting $\ip = 0$ we find a BPS solution for a real Higgs field
with the same magnetic field, and with the same Higgs VEV as the
real part (after the $U(1)_R$ rotation) of the original Higgs field.

The opposite direction is slightly more complicated. Let us assume that we
are given a solution of the BPS condition for a real Higgs field. We will show
that we can uniquely generate from it a solution for a complex Higgs field,
with any expectation value 
in the Cartan subalgebra for the imaginary part of the
Higgs field, perpendicular to the direction of the magnetic field.
The last requirement is necessary to ensure that $Q_M > 0$. 
We need to generate
an appropriate solution to equation \bps, 
i.e. define a field $\ip$ which equals
the desired VEV far along the $z$ axis (where we determined the asymptotic
values) and satisfies
\eqn\forip{D_i \ip = [\rp, \ip] = 0.}
The equation $D_i \ip = \del_i \ip + [A_i, \ip] = 0$ is of first order, and
we can easily solve it for a given field $A_i$. 
The solution, given the value of $\ip$ at a
point $x$ (which we choose far along the $z$ axis), is given by
\eqn\solip{\ip(y) = g_{xy} \ip(x) g_{xy}^{-1}}
where
\eqn\defgxy{g_{xy} = P \exp(-\int^x_y dx^i A_i).}
We should now show that the solution \solip\ is uniquely defined (i.e. it
does not depend on the path from $x$ to $y$), and that it satisfies $[\rp,
\ip] = 0$ everywhere. Both properties can easily be shown to hold as a
result of
the equality $[B_i, \ip] = 0$, which is satisfied by any solution
to $D_i \ip = 0$. First, any path which differs from the original path by a
small loop in the definition of $g_{xy}$ can
easily be shown to give a solution differing by $[B_i,\ip] = 0$. Second, we
chose $\ip$ so that at the point $x$ $[\rp,\ip]=0$, and then the equations
\bps\ lead to $D_i[\rp,\ip]=[B_i,\ip]=0$. Therefore, any solution to $D_i
\ip=0$ will automatically satisfy $[\rp,\ip]=0$ everywhere. 

Thus, up to a
choice of the VEV of the imaginary part of the Higgs field, there is a
one-to-one correspondence between classical BPS monopole solutions for
real Higgs fields and the classical BPS monopole solutions of the $N=2$
theory. The former were analyzed by Weinberg \weinberg, who found that, after
eliminating the Weyl transformation freedom as described above, the
dimension of the moduli
space of solutions with magnetic charges $n_m^i$ (for a given Higgs VEV)
is 
\eqn\nzm{4\sum_{i=1}^r n_m^i.}
For example, for the embedding of the 't~Hooft
Polyakov monopole in the direction of a simple root of the dual gauge group,
we find here just the three translational zero modes and the electric
charge zero mode transforming the monopole into a dyon.

By the discussion above, the same result 
applies also to the $N=2$ theory, after we
use the $U(1)_R$ freedom to set $Q_M > 0$ and eliminate the Weyl freedom by
choosing ${\rm Re}({\vec{h}}) \cdot \ai \geq 0$. However, it is important
to notice that for a particular $\vp$, we might need to perform different
Weyl transformations for different monopoles in the spectrum of the theory,
since we use different $U(1)_R$ transformations to set $Q_M > 0$.
For complex $\vp$ there is generally 
no choice of simple roots for which the number
of zero modes is given by \nzm\ for all monopoles, and we
need to be careful when computing the number of zero modes of several
monopoles for the same $\vp$ (i.e. at the same point in moduli space).

The correspondence we found between real and complex $\vp$ seems strange 
if we recall that Weinberg
interpreted \weinberg\ this moduli
space (for real Higgs field) by assuming the existence of fundamental
monopoles with charges $\bi$ ($i=1,\cdots,r$)
which have no force between them, so that
we can take them as far apart as we wish and still satisfy the BPS bound.
Equation \nzm\ then means that all
monopoles of higher charges can be interpreted as bound states of
the monopoles with charge $\bi$.
In the $N=2$ theory this is no longer the case. There is no longer any way
(for generic Higgs VEVs) to patch together monopole solutions in different
directions (in the Cartan subalgebra), and most of 
the formerly marginally stable bound state
solutions are now absolutely stable. However, we have assumed throughout
the discussion a localized solution, with a $1/r^2$ falloff of the magnetic
field. Clearly, 
a solution which consists of two monopoles
far apart does not satisfy this conditions, and, therefore,
cannot be generalized to complex fields as described above. 
Thus, there is no contradiction.
Still, it would be interesting to understand how to interpret
these classical ``multi-monopole'' solutions in the $N=2$ theory. 

There is a small subtlety which we have not yet addressed.
Equation \nzm\ for a real Higgs field is only correct when the real Higgs
field completely breaks the gauge symmetry to $U(1)^r$. We assume
throughout this paper that we are at a point at which the complex Higgs
field completely breaks the gauge symmetry. However, on special
surfaces in moduli space, which include the surface $\cm$ we were working on in
the previous section, it could happen that the real part of the Higgs
field (obtained after the $U(1)_R$ rotation as described above) does not by
itself completely break the symmetry. This happens if, after the $U(1)_R$
rotation, ${\rm Re}(\vp) \cdot \vec{\alpha} = 0$ for some root $\vec{\alpha}$.
Our correspondence between the
solutions for real and complex Higgs fields still works, but we must now
look at the computation for a real Higgs field which does not completely
break the gauge group. When the magnetic field has no component in the
non-abelian part of the remaining gauge group, this computation was also
performed by Weinberg in \weinbergg, with a result analogous to \nzm.
However, in the other cases (which include the surface $\cm$)
the number of normalizable bosonic zero modes
appears not to be completely known, and, therefore,
we do not know the number of zero
modes also in the $N=2$ theory in these cases. Since for this to occur we
need the projection of the Higgs field in the direction of two different
roots to be real, this problem may arise only on
singular surfaces on which states may be only marginally stable.
Our discussion in the following sections will therefore be
limited to generic points in the moduli space, where \nzm\ is indeed
correct.

\newsec{The low-lying semi-classical monopoles}

A general classical monopole solution
has a moduli space of bosonic zero modes, and in the semi-classical
computation we quantize the bosonic zero modes in this space. In general,
this moduli space is a complicated hyper-K\"ahler surface \atiyah, and the
quantization of the bosonic zero modes is difficult, even for $SU(2)$ as in
\suspect. There are $4$ bosonic zero modes which always exist,
corresponding to translations ($3$ zero modes) and to time-dependent gauge
transformations transforming the monopole into a dyon. When these are the
only zero modes that exist, as for the $n_m=1$ monopole for gauge group
$SU(2)$, the semi-classical
quantization of the bosonic zero modes is trivial, leading to
a series of dyon states whose electric charges differ by multiples of the
magnetic charge. Therefore, we start by checking when such a simple
moduli space occurs.

From the analysis of the previous section, it is clear that such a moduli
space may arise if, after the appropriate Weyl transformation, the
magnetic charge of the monopole is exactly a simple root of the dual gauge
group, $\vec{g} = \bi$, so that the vector $n_m^i$ is of the form
$(0,0,\cdots,0,1,0,\cdots,0)$. If $\vp$ is real (up to a global phase), we
use the same Weyl transformation for all monopoles, 
and then the monopoles corresponding to
simple roots indeed have a four-dimensional moduli space. 
All other monopoles for which all $n_m^i$ are positive
have a larger moduli space. For complex $\vp$ the situation is more
complicated, and generally monopoles corresponding to any root of the dual
group may be transformed by the Weyl transformation described above to
simple roots, in which case their moduli space would be simple. 
A particular case
which is easy to analyze is when $\vp$ lies in the surface $\cm$, defined
by $u_i=0$ for
$i=2,\cdots,N_c-1$. In this case, for $N_c \geq 3$, we can check and
see that no monopoles, including the simple root monopoles, are transformed
by the relevant Weyl transformations to simple root monopoles.
Thus, on this surface all monopoles have
a moduli space of dimension larger than $4$, whose quantization is complicated.

For general magnetic charges $n_m^i$ (which may be positive or negative), 
equation \nzm, which determines the number of zero
modes, is not necessarily positive. When the number is positive,
the analysis shows that if the monopole exists, the dimension of its
classical moduli space is \nzm, but it does not show that a solution indeed
exists. The BPS mass formula \bpsbound\ shows that 
for generic magnetic charges and at a
generic point in moduli space, if a solution exists it is stable.
The only generally known solutions when the gauge symmetry is broken to
the abelian subgroup are the embeddings of the
$SU(2)$ 't~Hooft 
Polyakov monopole, which give monopoles whose magnetic charge is a
root of the dual gauge group, and for these equation \nzm\ is always
a positive number. However, when \nzm\ turns out to be negative or zero, it
is obvious that no classical monopole solutions exist with these magnetic
charges, since any solution must have at least the translation zero modes.
For real $\vp$, when no Weyl transformations are necessary in the analysis
of the previous section, we can easily show that for every monopole whose
charges $n_m^i$ are not all non-negative (or all non-positive) there exist
regions in the moduli space for which this monopole has a non-positive
number of zero modes and, therefore, semi-classically it does not exist.

However, when we cross singular surfaces in moduli space, the number of
zero modes
\nzm\ may change for a monopole which is only marginally stable on the 
surface, and classical monopole solutions which did not previously
exist may arise. This phenomenon could occur also in the quantum theory
\jumps. Thus, monopoles having some positive and some negative $n_m^i$'s
may exist in parts of the moduli space. Such monopoles may also have
simple moduli spaces, like the simple root monopoles. For instance, in the
$SU(3)$ theory, if a monopole of charges $n_m^i=(2,-1)$ 
exists near the surface of real
$\vp$ it has just $4$ bosonic zero modes.

In fact, monopoles of this type must indeed exist in parts of the moduli
space if the transformation \btrans\ is indeed a symmetry of the theory.
This transformation takes quarks
whose electric charges are $\vec{e}_i= 
\vec{\mu}_i-\vec{\mu}_{i-1}$ (for $i=1,\cdots,N_c$,
where we define $\vec{\mu}_0=\vec{\mu}_{N_c}=0$), 
which always exist in the theory for weak
coupling, 
to states whose magnetic charges are proportional to $\vec{e}_i$.
For these magnetic charges (for
$i=2,\cdots,N_c-1$) we can see, by the analysis of the previous section, 
that there exist parts of the moduli space (for instance, near the surface 
of real $\vp$)
for which the computation gives a non-positive number of zero modes. 
Thus, for weak coupling these monopoles cannot exist there. 
For example, in the $SU(3)$ theory one of the quarks
has an electric charge $\vec{\mu}_2-\vec{\mu}_1$, which is transformed by
\btrans\ to magnetic charges $(1,-1)$. For these charges 
no solutions exist near
real $\vp$, when no Weyl transformations are involved in the analysis as
described in the
previous section. Obviously, if no classical solutions exist for these
quantum numbers we do not find any such states in the spectrum by the
semi-classical quantization.
However, near the surface $\cm$, defined by $u_i=0$
($i=2,\cdots,N_c-1$), for which the spectrum should be self-dual, at least
for $N_c=3$, we
find that these monopoles do always have a positive number of zero modes.
They may, therefore, exist classically, though the corresponding classical
solutions have not yet been built as far as we know. Thus, the classical
analysis does not rule out the existence of these states in part of the
moduli space (for weak coupling). 
It is easy to see that the monopoles related by \btrans\ to
the $W$ bosons have a positive number of zero modes at any point in moduli
space, since their
magnetic charge is proportional to a root of the dual gauge group (recall
that the roots transform among themselves under Weyl transformations).
In fact, the number of bosonic 
zero modes we find for these monopoles is always at least $4N_c$.

Next, we would like to discuss the fermionic zero modes in the background
of the monopole \zm. For any monopole, the gluinos have zero modes related by
supersymmetry to the bosonic zero modes, which restore the $N=2$
supersymmetry. In the absence of other
fermionic zero modes, these zero modes turn the monopole into an 
$N=2$ hypermultiplet \gauntlett. The
analysis of the quark zero modes is more complicated. Generally each quark
flavor may have $k$ zero modes in the background of the monopole field. For
$SU(N_c)$ gauge groups we can combine the zero modes of the quark and the
anti-quark into complex zero modes $\rho_{iA}$
($i=1,\cdots,N_f;A=1,\cdots,k$), which satisfy the anti-commutation
relations $\{ \rho_{iA}^{\dagger}, \rho_{jB} \} = \delta_{ij} \delta_{AB}$.
The zero modes $\rho_{iA}$ carry the same flavor quantum numbers as the
quarks. They are in a
fundamental representation of $SU(N_f)$ and carry a baryon number charge
equal to that of the quarks. 
By acting with the zero modes on the monopole
vacuum, we generate states in various (anti-symmetric) representations
of the $SU(N_f)$ flavor group. However, in general, not all these states
are BPS states. When the moduli space of the bosonic zero modes is
non-trivial, only a small number of these states actually become BPS-saturated
states when performing the semi-classical quantization. This is
demonstrated in the analysis performed for
$n_m=2$ and gauge group $SU(2)$ in \suspect. 

The number of zero modes of a quark in the
fundamental (or anti-fundamental) representation 
in the background of a monopole of
charges $n_m^i$ may be determined by an index theorem derived by
Callias \callias. 
This is generally a laborious calculation. However, for the 't~Hooft 
Polyakov embeddings, the problem is essentially reduced to an $SU(2)$
problem, and we may use the known results for $SU(2)$, computed
explicitly in \callias. Note that for these monopoles
a four-dimensional bosonic moduli space may arise, as discussed
above, in which case 
all fermion zero modes indeed generate BPS saturated states.
For these monopoles, the non-constant elements of the gauge fields $A_i^a$
and of $\Phi^a$ are in a $2\times 2$ block 
matrix, and thus only two elements of
the fermion (regarded as a vector in the fundamental of $SU(N_c)$) can be
non-zero. The VEV of the Higgs field in this $2\times 2$ matrix may be
divided into a part proportional to $\sigma^3$, which acts as an $SU(2)$
Higgs VEV, and a part proportional to the identity matrix, which
acts essentially as a (complex) mass term for the fermion. The analysis of
\callias\ was performed with a real mass term, and it
shows that a quark in the background of an $SU(2)$ monopole of
$n_m=1$, which is the monopole we are embedding, has one zero mode if the
mass $m$ lies between $a$ and $-a$ (where the Higgs VEV is $\pmatrix{a & 0 \cr
0 & -a \cr}$), and no zero modes otherwise. When there is an imaginary part
to the mass term, we should not actually look at zero modes of the Dirac
equation, since the BPS formula in the presence of a mass shows (in the
semi-classical limit) that
the mass of the BPS saturated state with the quantum numbers of the
monopole and fermion put together differs by ${\rm Im}(m)$ from that of the
monopole. Thus, in this case we should look at solutions to the Dirac
equation whose energy is ${\rm Im}(m)$, but this is exactly the energy that we
naturally find for the zero-mode solution of real mass in the presence of the
complex mass term.
Hence, the imaginary part of the mass term does not affect the analysis, and
we find that a quark has a zero mode in the background
of a 't~Hooft Polyakov embedding with magnetic charge
$\vec{g}=\vec{\beta}^i+\vec{\beta}^{i+1}+\cdots+\vec{\beta}^j$ if (and only if)
\eqn\zerom{|{\rm Re}(\vp) \cdot (\vec{\mu}_{i} - \vec{\mu}_{i-1} - \vec{\mu}_j
+ \vec{\mu}_{j+1})| \leq 
|\vp \cdot \vec{g}|} 
(where we choose the phase of
$\vp$ so that $\vp \cdot \vec{g}$ is real). 

It is easy to show that for real (or almost real) $\vp$ exactly one of the 
monopoles of charge $\bi$, which have a simple bosonic moduli space,
has a zero mode.
For general complex $\vp$ more complicated situations may arise, and in
particular near the surface $\cm$ it may easily
be seen that all the monopoles corresponding to roots of the dual gauge
group (i.e. the 't~Hooft Polyakov embeddings) have quark zero modes.

\newsec{Duality of the spectrum of $N=2$ SQCD}

We would like now to bring together all of our results, and see
if we can find a consistent electric-magnetic duality transformation for
the $N=2$ SQCD theory. For all electric and magnetic charges, we should find
the same flavor quantum numbers for the
states which transform into each other under the duality, up to a possible
automorphism of the $SU(2N_c)\times U(1)_B$ flavor group. 

In order to compare states related by the duality we should generally
compare states at weak coupling with states at strong coupling. 
However, since our
methods for computing the particle 
spectrum are semi-classical, we can only compute
the spectrum at weak coupling. Thus, such a comparison is generally
impossible, unless we can connect the spectrum at weak coupling with the
spectrum at strong coupling. For $G=SU(2)$, for the $N=4$ SYM theories 
and for $G=SU(3)$ along the
surface $\cm$, we can see from \ad\ 
that we do not cross any singular surfaces
in moving from strong to weak coupling.
Hence, the particle spectrum cannot change. Along
this surface, which we assume to be transformed to itself by the duality,
the spectrum found at weak coupling by the semi-classical computation
should be self-dual. In other parts of the moduli space, we cannot
discard the possibility
that the spectrum changes at some value of $\tau$, and we
cannot generally make such a comparison. In fact, we will see that in
other parts of the moduli space the semi-classical spectrum is not
self-dual, and such
phase transitions changing the spectrum of the theory apparently 
do indeed occur as we change the value of $\tau$.

There are some states of the theory which always exist in the
semi-classical region. These are the quark hypermultiplets
in the fundamental of $SU(N_f)$ with baryon number $B=1$, and the $W$ 
bosons which reside in vector multiplets and are singlets of the flavor group.
These representations should be identical to those we find for the monopole
states connected with the quark and $W$-boson 
states by duality, up to a possible automorphism of the flavor group.

As discussed in the previous section, the monopole generally gets flavor
quantum numbers due to the zero modes of the quarks, which exist in the
background of the monopole \zm, acting on the monopole state.
Acting with the zero modes may also generally change the electric charge of
the monopole state.
For $SU(2)$, there exists a simple projection determining
which flavor states may actually have $n_e^i=0$, given by 
equation (5.2) of \swii. This equation
relates the charge of the state under the center of the gauge group,
which is naturally determined by the electric
charge of the state, with the charge of the state under the center of the
flavor group. However, we do not know how to generalize this
equation to a general gauge group, since the gauge transformation used
to derive this equation (a rotation around the direction of the Higgs
field \dyons) is generically not in the center of the gauge group.

In the $SU(2)$ case, Seiberg
and Witten argue (at the end of section 5 of \swii) that
the center of the flavor group is always faithfully represented on the
lattice of charges, i.e. that the charge of a state uniquely determines the
charge of the flavor representation of that state under the center of the
flavor group. However, it is not clear whether this should be true in general.
In fact, we can easily see that it is not true even for the elementary
quark states in the $SU(N_c)$ case. The center of the flavor group for $N_c
> 2$ includes a $\bZ_{2N_c}$ group, and the sum of the electric charges of
$N_c-1$ 
different quark states (whose $\bZ_{2N_c}$ flavor
group charge is $1$) gives the electric
charge of an anti-quark state (whose $\bZ_{2N_c}$ charge is $-1$). 
However, for the
elementary electric states the gauge group $\bZ_{N_c}$ charge 
equals (modulo $N_c$) the flavor group $\bZ_{2N_c}$ charge.
Therefore, it seems possible to have a relation between
the flavor $n$-ality of a state (modulo $N_c$) and its electric charge.
This relation certainly exists for the elementary electric
states. It is satisfied also by the monopole states if the color
$n$-ality of the monopole vacuum, on which the quark zero modes act, is
zero, which seems to be the case. 

The electric charge of the quark 
zero modes is not well defined in general, but
their $n$-ality (the $\bZ_{N_c}$ charge) is well defined.
Thus, we expect that states generated by acting with $k$ quark zero
modes on the monopole vacuum
will have an $n$-ality which is larger by $k$ than that of the
monopole vacuum. For a 't~Hooft Polyakov embedding of magnetic charge
$\vec{g}=\vec{\beta}^i+\vec{\beta}^{i+1}+\cdots+\vec{\beta}^j$, 
the only non-zero
color components of the quark zero modes have electric charges
$\vec{\mu}_i-\vec{\mu}_{i-1}$ and $\vec{\mu}_{j+1}-\vec{\mu}_j$.
We expect, therefore, the states generated by the zero modes
to differ by these charge vectors. For these monopoles with magnetic charge
${\vec{g}}$ it does not matter 
which of the two charge vectors
we take, since the difference between them is $\vec{g}$, and the
$T^2$ transformation relates states differing by electric charge
$\vec{g}$.

Let us now discuss how all this
affects the duality of the spectrum. We will discuss in detail only 
the gauge group $SU(3)$ case, though most of the discussion may be easily 
generalized. As discussed above, the only region where we can easily see
that the particle
spectrum must be self-dual is
near the surface $\cm$ with $u_2=0$. Along this
surface, to the extent that we have computed it, we find that the 
transformation \btrans\ is consistent with the
semi-classical spectrum. All of the magnetic charges
related to quarks and $W$ bosons by this transformation give a positive
number of zero modes by \nzm, including the $n_m^i=(1,-1)$ monopole. The
resulting monopoles all 
have a non-trivial monopole moduli space (with at least
$12$ bosonic zero modes), and may also have
fermionic zero modes that would give precisely the correct flavor quantum
numbers for these states to be related to the electric states. 
The 't~Hooft Polyakov embeddings also have
complicated moduli spaces (with $8$ bosonic zero modes)
in the vicinity of $\cm$, and for all of them the quarks have zero
modes. Since the bosonic moduli space is complicated for all of the
monopoles, it is not clear which flavor representations actually
exist as BPS saturated states generated from these monopoles, 
and at what electric charges. 
The ``electric-magnetic'' duality of \btrans\
transforms some of these states among themselves, and looks like it could be
consistent with the semi-classical spectrum. 
A more comprehensive analysis is necessary,
however, to verify that (at least near the surface $\cm$) 
the spectrum is indeed self-dual. 

Next, let us analyze the semi-classical particle spectrum for real 
(or close to real) $\vp$
($4u_2^3/27u_3^2 > 1$ in gauge invariant variables), and check if it is
self-dual. 
In this case, as discussed above, both monopoles corresponding to the
simple roots $\bi$ of the dual gauge group have a simple bosonic moduli
space, and one of them has a fermionic zero mode while the other does not.
For the monopole which has a fermionic zero mode, we act with the quark
zero modes to get states in the 
$\bf{1+6+15+20+{\overline{15}}+{\overline{6}}+1}$ 
representations of the $SU(6)$ flavor group, with rising
baryon numbers from left to right. It is not clear how to determine the
absolute baryon number of the monopole state, because we can always add to
the baryon number any linear combination of conserved charges.
Therefore, we will ignore the baryon number in our
analysis. If this monopole is related by duality to a quark state (as for
$SU(2)$), we expect to find 
just a $\bf{6}$ representation (or a $\bf{\overline{6}}$ if there is an
outer automorphism of the $SU(6)$ flavor group involved).
In section 3 we showed that states related by duality to 
quarks always have an electric
charge vector proportional to the magnetic charge vector. 
For $\vec{g}$ corresponding to a root vector, this means
that their electric charge must be a multiple of $\vec{g}$, which must be
zero
(up to $T^2$ transformations) for it to lie in the electric charge lattice.
It is not apriori clear which of the states generated
by the quark zero modes has zero electric charge, since we do not know the
electric charge of the monopole vacuum. However, at least for real $\vp$
there seems to be
(classically) a CP symmetry in this
theory (as described in \ms\ for the $SU(2)$ case)
inverting the electric charges but not the magnetic charges (for
$\theta=0$). In this case, obviously, the state in the
$\bf{20}$ representation has zero electric charge. Thus, if the
semi-classical analysis is relevant in this part of moduli space, then this
monopole
cannot be related to a quark by duality. In fact, by similar arguments we
can show that the monopoles generated by 't~Hooft Polyakov embeddings,
whose magnetic charges are roots of the dual gauge group, can never be related
to the quarks by duality for $N_c > 2$.

The monopole corresponding to the other simple root 
has no quark zero modes in this case, and it is, therefore, a
flavor singlet which resides (after using the gluino zero modes) in an $N=2$
hypermultiplet. There is no similar state with only electric charges. Hence,
this monopole also cannot transform into purely electric
states. In any case it seems that, if the semi-classical analysis is relevant,
the quarks are not connected to the ``fundamental''
monopoles of charge $\bi$ by a duality transformation. 

The quarks could still be related to
monopoles of higher magnetic charge, as in \btrans. 
The semi-classical quantization of
these monopoles is much more difficult.
However, as discussed in the previous section,
this transformation takes the quark state with electric charge
$\vec{e}=\vec{\mu}_2-\vec{\mu}_1$ to a monopole of charges $n_m^i=(1,-1)$
which does not exist near
the surface of real $\vp$, and obviously it cannot be the correct
transformation there.
Thus, it seems that the semi-classical particle spectrum near the 
surface of real $\vp$ is not self-dual. 
As discussed above, in this area of moduli
space we cannot rule out changes in the spectrum as we go from strong to
weak coupling, and the semi-classical spectrum does not have to be
self-dual for the duality to hold. Our discussion shows that if the duality
holds, then the spectrum necessarily changes as we move from strong to
weak coupling near the surface of real $\vp$. 

The situation in all $SU(N_c)$ gauge theories
is actually similar to the situation
described above for the $SU(3)$ theory.
For all $N_c$, we find
near the surface $\cm$ that all of the monopoles related by \btrans\ to
the quarks may indeed exist, and always have many bosonic zero modes, making
their semi-classical quantization complicated. The 't~Hooft
Polyakov embeddings all have at least $8$ bosonic zero modes near this
surface. Thus, we cannot even tell which BPS states are generated from
these monopoles 
near this surface without performing complicated computations.
The duality seems to be consistent along the surface $\cm$, even though we
have not been able to show that this must be the case for $N_c > 3$.
However, many possible
transformations are consistent with the spectrum at the classical level,
and it appears to be necessary to perform (at least)
the full semi-classical calculations (as in \suspect)
in order to really check the duality. For all $SU(N_c)$ groups ($N_c > 2$)
we find that the particle spectrum for real $\vp$ is not self-dual.
Therefore,
if the duality holds, the spectrum in this region must change as we move
between strong and weak coupling.

\newsec{$N=2$ SQCD with non-zero quark masses}

In our discussion up to now we focused on the case of zero quark masses. In
fact, it seems that all tests of $S$ duality so far, except for the duality
of the low energy spectrum-generating curve, have focussed on the massless
case (in the $N=4$ theory they were performed without a mass term for the
adjoint field which breaks the supersymmetry to $N=2$). However, apriori
we see no
reason why exact electric-magnetic duality may not also be true in theories
with quark masses which transform appropriately under the duality.

Thus, we would like to try and test the duality by semi-classical
computations also in the theories with massive quarks. 
Note that mass terms explicitly break the scale invariance and $U(1)_R$
symmetries that we used in the classical analysis of sections 4 and 5, and
the analysis has to be changed appropriately.
However, this analysis runs
into the same problems that we ran into in the previous sections for cases
other than $G=SU(3)$ along 
the surface $\cm$.
Again, we cannot prove that when we go from strong to weak coupling we do
not cross singular surfaces, changing the particle spectrum of the theory.
In fact, this is true even for the $SU(2)$ theory of \swii. For instance,
if we look at this theory for $4$ equal quark masses $m$ and for weak
coupling\foot{We thank M. R. Plesser for this
example.}, 
then we know that for $u \ll m^2$ the theory looks like an $N_f=0$
theory, with a dynamically generated scale $\Lambda$ proportional to $m
q^{1/4}$ (where $q = e^{i\pi \tau}$).
In this theory we know \swi\ that there are two singularities, at $u =
\pm \Lambda^2$, there is a marginal stability curve passing through 
both of them, and
the particle spectrum of the theory changes when we cross it. Since
$\Lambda$ depends on $\tau$, it is clear that this curve moves as we change
$\tau$, so that the spectrum for the same $u$ at strong and weak coupling
is not necessarily the same. 

As in the previous section, we can see that the semi-classical spectrum of
this theory is not invariant under the $S$ duality transformation of \swii,
signifying that such a change in the spectrum does indeed occur. The $S$
duality of \swii\ transforms the theory with masses $m_i=(m,m,m,m)$ to a
theory with masses $m_i=(2m,0,0,0)$. The $SU(4)$ symmetry acting on the
four quarks is transformed to the $SO(6) \sim SU(4)$ symmetry acting on the
last three massless quarks. The $SU(4)$ quantum numbers of states
related by the duality should be the same. In the theory with
masses $m_i=(m,m,m,m)$, when $0 < u < m^2$, 
the quarks have no zero modes in the
background of the $n_m=1$ monopole.
Therefore, semi-classically
there are no states in the $\bf 6$ representation of the $SU(4)$
flavor group, because these may only be generated by quark zero modes.
However, in
the theory with masses $m_i=(2m,0,0,0)$, the three massless quarks are in
the $\bf 6$ representation of $SO(6) \sim SU(4)$, and these obviously exist
at weak coupling for all $u$. Thus, at least in part of the moduli space,
the spectrum must indeed change from strong to weak coupling for the
duality to hold.

This phenomenon is completely general. For
instance, for general $N_c$, a quark with mass $m_i$ becomes marginally 
stable when ${\rm Im}({{a_1 + m_i/\sqrt{2}}\over a_D^1}) = 0$. The form
of this curve in moduli space changes when we change the coupling $\tau$,
on which both $a$ and $a_D$ depend since they are periods of a curve depending
on $\tau$. This is true also for the $N=2$ theory with a massive
hypermultiplet, obtained by a mass perturbation from the $N=4$ theory.

Therefore, in general, we cannot check the duality in these theories
by requiring that the semi-classical spectrum should be self-dual, up to
the transformation of the masses. For small masses (relative to $\vp$), 
it is clear, since the
curves all depend continuously on the masses, that near $\cm$ the spectrum
should still be self-dual, since, in this case, 
we do not expect to cross surfaces along
which the low-lying states are marginally stable.
The monopole spectrum is self-dual in this case just as in the zero
mass case, since when the masses are small the fermions
still have the same zero modes. The quark-mass corrections to the BPS
formula for the monopole mass are negligible for small masses at weak 
coupling. However,
without computing the explicit form of the singular surfaces, we cannot
check the duality by semi-classical computations in the general
massive case.

\newsec{Summary and conclusions}

In this paper we analyzed $N=2$ SQCD theories with zero beta function, in
an attempt to understand their duality transformations and check them by a
semi-classical analysis. We found that apriori many transformations in the
$Sp(2r,\bZ)$ low-energy symmetry group may be exact symmetries of the theory.
Unlike the $N=4$ and $SU(2)$ cases, the transformation
$\tau \to -1/\tau$ is not a symmetry in the general case.
Clearly, more constraints are needed to find what is the exact duality
group of the theory. 
Unfortunately, the semi-classical analysis does not
provide many constraints on the duality
in these theories, since in general the particle
spectrum at weak coupling and at strong coupling need not be the same. The only
part of moduli space for which the spectra must be the same is the
surface $\cm$ for the case of $G=SU(3)$ with
zero quark masses. In this case we found
that the computations involved in the semi-classical quantization are
difficult even for the monopoles of small $n_m^i$, and we did not
perform them. The only test we can do without performing either the
computations on the moduli space of these monopoles, or the computation of
the singular surfaces from the spectrum-generating curve, is to check that
classical solutions may exist with appropriate magnetic charges to be related
by duality to the quarks and $W$ bosons. This is indeed the case, though
these classical solutions have not yet been constructed. For higher
$SU(N_c)$ groups we also found the duality to be consistent with our
classical computations along the surface $\cm$. Quantum corrections appear
to be important for $N_c > 3$, as discussed at the end of section 3,
and these theories deserve further investigations.

Our discussion was limited to $SU(N_c)$ gauge groups, but all of it can be
straightforwardly generalized to any gauge group. For other gauge
groups the calculation of section 3 will give different constraints, and
perhaps simple semi-classical computations may give more constraints on the
duality there, provided that the analysis of the particle spectrum 
along the surface analogous to $\cm$ is simpler.

Obviously, more tests should be made to verify the existence of an
exact electric-magnetic duality transformation in these theories, and to
find its exact form. One possible direction is to make better
semi-classical computations, by generalizing the computations of
\suspect\ to the monopole moduli spaces which we find in these theories.
Another direction is to compute
the singular surfaces of these theories, in order to find the exact
limits of validity of the semi-classical analysis, i.e. the values of
the $u_i$ for which a certain state does not become marginally stable when
going from weak to strong coupling. Other directions for verifying the
duality may involve computing
various partition functions of these theories, perhaps after twisting them
to topological theories.

\centerline{ }
\centerline{\bf Acknowledgments}

We would like to thank C. Imbimbo, Y. Oz, M. R. Plesser, N. Seiberg and 
J. Sonnenschein for useful discussions on this subject.

\listrefs

\end